# Fast and Accurate Optical Fiber Channel Modeling using Generative Adversarial Network

Hang Yang, Zekun Niu, Shilin Xiao*, Jiafei Fang, Zhiyang Liu, David Fainsin and Lilin Yi

*Abstract*—In this work, a new data-driven fiber channel modeling method, generative adversarial network (GAN) is investigated to learn the distribution of fiber channel transfer function. Our investigation focuses on joint channel effects of attenuation, chromic dispersion, self-phase modulation (SPM), and amplified spontaneous emission (ASE) noise. To achieve the success of GAN for channel modeling, we modify the loss function, design the condition vector of input and address the mode collapse for the long-haul transmission. The effective architecture, parameters, and training skills of GAN are also displayed in the paper. The results show that the proposed method can learn the accurate transfer function of the fiber channel. The transmission distance of modeling can be up to 1000 km and can be extended to arbitrary distance theoretically. Moreover, GAN shows robust generalization abilities under different optical launch powers, modulation formats, and input signal distributions. Comparing the complexity of GAN with the split-step Fourier method (SSFM), the total multiplication number is only 2% of SSFM and the running time is less than 0.1 seconds for 1000-km transmission, versus 400 seconds using the SSFM under the same hardware and software conditions, which highlights the remarkable reduction in complexity of the fiber channel modeling.

## I. INTRODUCTION

THE modeling of optical fiber channel is significant for system designs and simulations. The conventional channel modeling is based on split-step Fourier method (SSFM), which is carried out by solving the nonlinear Schrödinger equation (NLSE) approximately [1]. However, the iteration steps of SSFM result in high complexity of computation. To avoid such computational complexity, many fiber channel models are proposed to estimate the channel conditions directly [2-5]. For example, the Gaussian noise (GN) model evaluates the signal-to-noise ratio (SNR) of fiber channel with minor inaccuracies, and the closed-form approximation of GN model further reduce the complexity and offer almost real time calculation for performance prediction [4, 5]. Although GN model class can predict the SNR of the signal accurately, they are not available to model the specific distortions during the fiber transmission. Therefore, the fast and accurate channel modeling to reflect the specific distortions is still an open issue.

Recently, deep learning (DL) is utilized as a data-driven method in channel modeling, which can fit the channel transfer functions by neural networks (NNs) according to the channel input and the output data [6]. Compared with the conventional model-driven method, the data-driven approach prevents complex mathematical theories and expert knowledge [6, 7]. Moreover, the calculation operation of DL is mainly composed of multiplications and additions without complicated operations, such as fast Fourier transform (FFT) in the SSFM-based modeling method. With the help of graphic processing units (GPU), the parallel calculations of NNs can be realized, which further improves the running speed of the model [8]. Additionally, the channel model based on DL has high compatibility with other neural network structures. For example, the DL-based channel model has been considered as an approach to address the gradient back-propagation problems of the end-to-end communication system by embedding the DL-based channel model into the autoencoder structure [9]. Moreover, the NN-based models can also be utilized for performance prediction. The noise power can be obtained by the NN model transmission and digital signal processing (DSP). However, the traditional NNs with mean square error (MSE) as loss function, such as back-propagation deep neural network (BP-DNN), only approximate the means of the distributions of the channel. These NNs are not able to model the optical fiber channel with the random distortions, such as the amplified spontaneous emission (ASE) noise.

As a deep generative model, a generative adversarial network (GAN) consists of two neural networks, which are trained by playing games against each other. GAN can learn and represent a random distribution, $p_{data}$, through a neural network [10], and the new samples satisfying the target distribution can be generated. In recent years, GAN has been widely used in image and visual computing, language and speech processing, etc. [11-15]. Considering the communication channel can also be regarded as a generative model that satisfies a certain

This paper is submitted for review on August 16, 2020 and is supported by National Key R&D Program of China (2018YFB1800904), National Nature Science Fund of China (No. 62071295, No. 61775137, No. 61431009, No. 61433009), and National "863" Hi-tech Project of China.

H. Yang, Z. Niu, S. Xiao, J, Fang, Z. Liu and L. Yi are with State Key Laboratory of Advanced Optical Communication System and Networks, Department of Electronic Engineering, Shanghai Jiao Tong University, Shanghai, 200240, China (e-mail: hangyang@sjtu.edu.cn, zekunniu@sjtu.edu.cn, slxiao@sjtu.edu.cn, jiafeifang@sjtu.edu.cn, dextermorgen@sjtu.edu.cn, lilinyi@sjtu.edu.cn).

D. Fainsin is with IMT Atlantique, 2 rue Alfred Kastler, 44307 Nantes, France (e-mail: david.fainsin@imt-atlantique.net).

conditional distribution $P(y|x)$, GAN is recently proposed to estimate the distribution of the channel transfer function and generate new channel output data with the same distribution [9, 16-18].

GAN was firstly proposed to model the addictive white Gaussian noise (AWGN) channel and Rayleigh fading channel in [9]. Timothy J. O'Shea used variational GAN to learn the accurate probability distribution function of the AWGN channel and nonlinear channel [16], which proved the ability of GAN to model the noise and nonlinear channel distribution. The authors of [17] utilized a new GAN structure to learn the inter-symbol interference (ISI), verifying that GAN can learn the memory effects by designing the condition vector. In the case of optical communication system modeling, the authors of [18] used GAN to model the actual optical intensity modulation/direct detection (IM/DD) system, but their work mainly described how GAN is applied to end-to-end communication systems. The performances and results of GAN have not been verified. Another NN structure, bi-directional long short-term memory (BiLSTM), was proposed for optical fiber channel modeling and demonstrated to have good fitting ability to learn the characteristics between data sequences, including dispersion and nonlinearity [6]. However, their work focused on short-distance channel modeling within 80 km and ignored the modeling of noise.

In this paper, we employ the GAN to model the optical fiber channel with the characteristics of chromatic dispersion (CD), self-phase modulation (SPM), attenuation, and amplified spontaneous emission (ASE) noise induced by erbium-doped fiber amplifier (EDFA). We modify the loss function, design the condition vector, and address the mode collapse of GAN to realize the modeling of the fiber channel. The mode collapse refers to the neglect of data diversity generation. The effective architecture, parameters and training skills of GAN are also shown in the paper. The ability of GAN to estimate the transfer function of the fiber channel is demonstrated from constellations, optical waveforms, spectra, and the normalized mean square errors (MSEs). The exact constellation shapes, high-overlap ratio of waveforms, and low normalized MSEs verify the high accuracy of GAN-based fiber channel modeling. Additionally, the GAN-based model is demonstrated to have good generalization abilities under different optical launch powers, modulation formats, and input signal distributions. Compared with the conventional SSFM-based fiber modeling method, the total multiplication number of GAN is only 2% of SSFM, and the running time is less than 0.1 seconds for 1000-km fiber transmission, while the running time is about 400 seconds by using the SSFM under the same hardware and software conditions. To the authors' best knowledge, the paper is a detailed study of GAN for dispersive and nonlinear fiber channel modeling at arbitrary distance and launch power firstly. The proposed GAN is a general method, which can be regarded as a powerful tool for devices and the whole communication systems modeling in the future.

The rest of the paper is organized as follows. Section II introduces the optical transmission system structure we used in this work. Section III introduces the principle of proposed optical fiber channel modeling, including the GAN working mechanism and implementation details during the training process. In Section IV, the results are presented and discussed in detail. The conclusions are drawn in Section V.

## II. OPTICAL FIBER COMMUNICATION SYSTEM STRUCTURE

As shown in Fig. 1, we simulate an optical fiber communication system to test the versatility of the GAN. The GAN is implemented to replace the fiber channel and EDFA to demonstrate the ability of modeling the joint effect of CD, SPM, attenuation, and ASE noise. The transmitter consists of modulation, oversampling, pulse shaping, and power normalization. We assume that the transmitter uses 16 quadrature amplitude modulation (QAM), and all the symbols and samples are expressed as complex values in this system. Following the four times up-sampling, root raised cosine (RRC) filter is used for signal shaping, which satisfies the Nyquist criterion and effectively avoids ISI [19]. Power normalization controls the transmission power of the optical signal, and then the signal is input to the standard single-mode fiber (SSMF).

The propagation of light in a lossless optical fiber can be governed by the nonlinear Schrödinger equation (NLSE) [20], which is simplified as

$$\frac{\partial A(z,t)}{\partial z} = \left( \hat{D} + \hat{N}\left[ A(z,t) \right] \right) A(z,t), \quad (1)$$

where $\hat{D}$ is the linear component, which denotes the effects of attenuation and CD, and $\hat{N}$ is the nonlinear component, representing the self-phase modulation effects related to the signal energy. $A$ denotes the optical field complex envelope, $z$ represents the distance, and $t$ is the time. Although NLSE provides the mathematical fundamentals for optical fiber channel modeling, there is no analytical solution. Thus, numerical approaches are used for simulation. The most commonly used scheme for solving (1) is SSFM [1], which can be expressed by

$$A(z+h,t) \approx \exp\left(\frac{h}{2}\hat{D}\right) \exp\left\{ h\hat{N}\left[ A(z+\frac{h}{2},t) \right] \right\} \exp\left(\frac{h}{2}\hat{D}\right). \quad (2)$$

After each span, an EDFA is used for amplification. However, EDFA introduces ASE noise simultaneously, whose power spectral density is

$$S_{ASE}(f) = h_p \nu n_{sp} \left[ G(f) - 1 \right], \quad (3)$$

where $h_p$ is Plank's constant, $\nu$ is the carrier frequency, $G$ is the amplification gain, and $n_{sp}$ is the spontaneous radiation factor related to the noise figure of EDFA [21]. The receiver uses a matched RRC filter and then performs digital backward propagation (DBP) algorithm, an inverse process of SSFM, to compensate CD and nonlinearity [22]. Then down-sampling and demodulation are performed subsequently. The main fiber channel parameters of this work are shown in TABLE I.

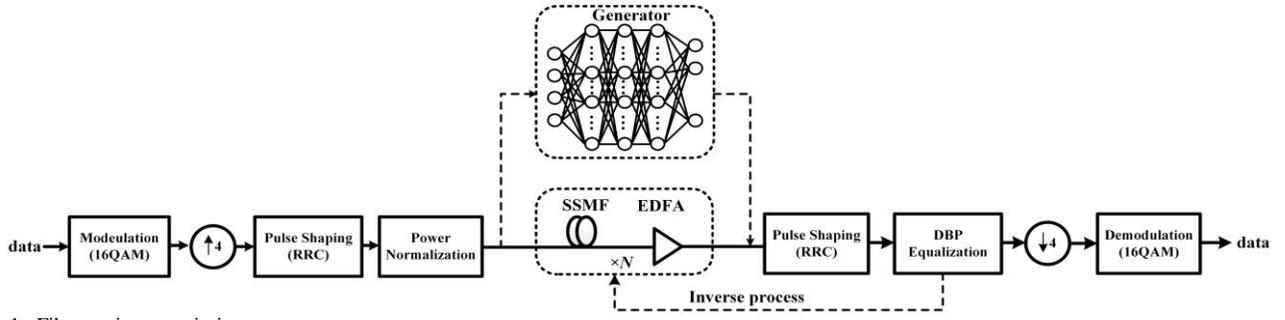

Fig. 1. Fiber-optic transmission system.

Considering that the channel transfer function has no analytical expression, we cannot directly compare the differences between the channel modeled by SSFM and GAN, which are represented by $f(\cdot)$ and $g(\cdot)$, respectively. Therefore, the DBP compensation is utilized as an auxiliary verification method, as in Fig. 2. DBP compensation can be regarded as the inverse function of the channel transfer function $f(\cdot)^{-1}$. If the DBP equalized output by GAN is similar to the channel input, i.e., $\hat{x}$ is close to $x$ in constellations, $g(\cdot)$ and $f(\cdot)^{-1}$ can be considered as a pair of inverse functions. Then $g(\cdot)$ and $f(\cdot)$ can be regarded as equivalent, proving that GAN can estimate the distribution of the optical fiber channel transfer function.

TABLE I
Fiber channel parameters

| Parameters | Value |
|---|---|
| Carrier length | 1550 nm |
| Symbol rate | 30G Baud |
| Attenuation | 0.2 dB/km |
| Dispersion | 16.75 ps/(nm·km) |
| Dispersion slop | 0.075 ps/(nm²·km) |
| Nonlinear refractive index | 2.6e-8 μm²/W |
| Core Area | 80 m² |
| Step distance | 0.01 km |
| Span length | 50 km |
| Noise figure | 5 dB |

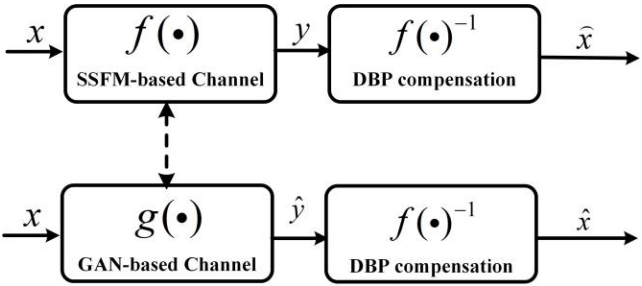

Fig. 2. The comparison scheme of GAN-based and SSFM-based channel modeling with DBP compensation.

III. PRINCIPLE OF PROPOSED OPTICAL FIBER CHANNEL MODELING

GAN is an example of generative models using an adversarial process, and it consists of two parts: generator and discriminator represented by a deep neural network (DNN), respectively. The authors of [23] proposed to add some extra conditional information to the GAN to control the generated outputs, which is called conditional GAN (CGAN). We adopt CGAN to model the optical fiber channel, and the structure of CGAN is shown in Fig. 3. Generator aims to capture the training data distribution and generate new data with the same distribution to fool the discriminator. A vector of noise $z$ with a prior distribution $p_z(z) \sim \mathcal{N}(0, I)$ and the condition vector $x$ are input into the generator and mapped to the generative fake data. Different noise vectors can map to different generative data. The condition vector $x$ defines the characteristics of the generative data. Discriminator classifies the real data and fake data with the addition of the condition vector $x$. The output of discriminator $D(x)$ represents the probability that $x$ is the real data. If $D(x)$ equals to 1, $x$ is determined to be true data. On the contrary, if $D(x)$ equals to 0, $x$ is determined to be fake data.

The generator and discriminator are trained alternately in an adversarial process to find a Nash equilibrium as the optimization. The total optimization loss function can be represented as

$$\min_G \max_D V(D,G) = \mathbb{E}_{y \sim p_{data}(y)}[\log D(y|x)] + \mathbb{E}_{z \sim p_z(z)}[\log(1 - D(G(z|x)))], \quad (4)$$

where $p_{data}(y)$ is the distribution of the real data and $p_z(z)$ is the distribution of the noise. $D(y|x)$ represents the discriminator output corresponding to the real data on the condition of $x$. $G(z|x)$ represents the generative data on the condition of $x$ [24]. The generator is trained to minimize $\log(1-D(G(z|x)))$, meaning that the generator tries to fool the discriminator to label the fake data approximate to 1. And the discriminator is trained to maximize $\log D(y|x)$ for the real data and $\log(1-D(G(z|x)))$ for the fake data, indicating it tries to label the real data to 1 and the fake data to 0. In the end, the ultimate goal is to make the discriminator unable to estimate whether the output of the generator is true, i.e. the output of discriminator approximating to 0.5.

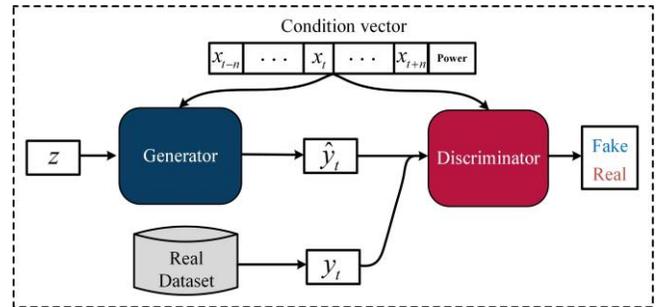

Fig. 3. Structure of GAN.

When the generator and discriminator fail to converge to the Nash equilibrium, a common problem, mode collapse, may be encountered [24]. Mode collapse problem refers to that generator only produces some modes of the data in the training set, and neglects other modes. Fig. 4 illustrates the mode collapse problem, where the blue line represents the probability density function (PDF) of the training data, and the red line represents the PDF of the generated data. There are three main modes of the target distribution. But the generated data by GAN only contains a single mode, which presents the diversities of the generated data are significantly reduced. By enhancing the ability of generator, such as improving network architecture and adopting the new optimization method, the mode collapse of GAN can be alleviated [25]. Experience shows that increasing training dataset is always useful to improve the learning ability of GAN.

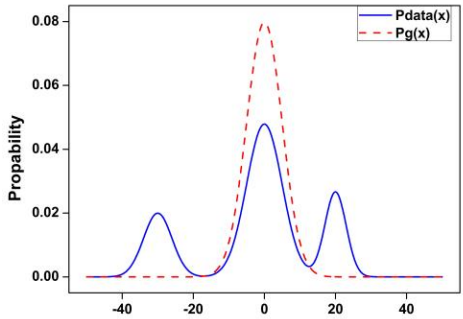

Fig. 4. An illustration of the mode collapse problem.

In this work, we collect the samples of optical transmission before and after the fiber as the condition vector and real data of GAN, respectively. When using the normal loss function as (4) in the training process, the output generated by GAN is not accurate compared with the output generated by SSFM. In the frequency domain, the large pulses exist in the spectra of GAN output. To improve the accuracy of channel modeling, we calculate the reconstruction error between real and fake data as a regularization term to the loss of the generator, which can be shown as

$$\mathrm{L}_y = \min_G \mathbb{E}_{x,y \sim p(x,y)} \| y - G(z|x) \|_1, \qquad (5)$$

where the generative data aim to be as close to the real data as possible. This modified loss function is called Ly loss in this article. It acts a restriction on the signal generation, which ensures that the amplitudes of the generated signal are in the same range as that of the real signal. Note that Ly loss does not impede the adversarial training process of generator and discriminator. We find that Ly loss can also accelerate the training process, which helps GAN converge to an equilibrium rapidly. Meanwhile, the Ly loss calculation will not cause any computational difficulties for the training. The effects of the modified loss function will be discussed in Section IV.

As shown in Fig. 5, the condition vector structure is customized for fiber channel modeling to improve modeling accuracy, flexibility, and training speed. Firstly, considering ISI caused by CD, the condition vector needs to contain both current and the surrounding samples, which can help GAN learn the relevance between sequences. The current time-step transmit samples are centered in the condition vector $x=[\,x_{t-n},\,...,\,x_t,\,...,\,x_{t+n}]$, where $x_t$ represents the current transmitted samples, $[x_{t-n},\,...,\,x_{t-1}]$ denotes the samples that have been transmitted before the current samples, and $[x_{t+1},\,...,\,x_{t+n}]$ are the future samples to be transmitted. The number of surrounding samples $n$ is proportional to the delay caused by CD, relating to the transmission length. By calculating the number of symbols affected by ISI and matching the transmission rate of the signal, $n$ is set to five for each span in this paper. Next, we find that four samples estimation, i.e., generating one symbol each time, is more accurate than one sample estimation. It is also feasible to generate more symbols each time, but the longer training time is required. Therefore, the generator is designed to generate four samples each time in this work. Moreover, considering that the input of the neural network must be a real number, we concatenate the real and imaginary part of a complex number to form a one-dimensional array, i.e., the in-phase and quadrature parts of the channel input are connected as shown in Fig. 5. In conclusion, the symbol $x_t$ is represented by eight real number, the length of the generator output is eight, and the number of surrounding samples for each span is 80 ($5 \times 8 \times 2$) in total. Finally, in the aspect of flexibility to transmit power, we add the value of optical launch power to the condition vector, whose unit is $mW$. In a word, the training dataset includes the input and output data of the fiber channel with different optical launch powers.

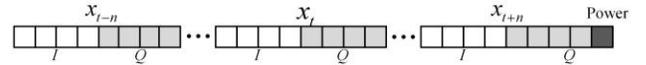

Fig. 5. Condition vector structure of GAN.

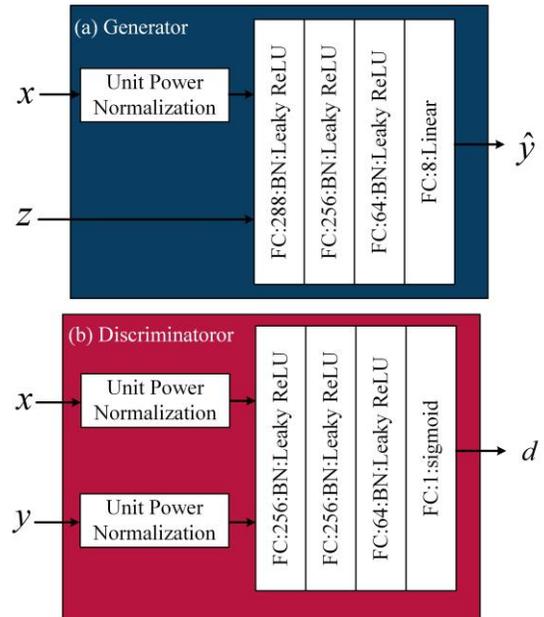

Fig. 6. A diagram of the architecture of Generator and Discriminator networks. The notation "FC:288:BN:Leaky ReLU" signifies that the layer contains 288 fully connected neurons with the batch normalization and a leaky Relu activation function.

The structures and parameters of the generator and discriminator are shown in Fig. 6. The channel input and output should be unit power normalized firstly before input to the generator and discriminator as

$$\tilde{x}_i = \frac{x_i}{\sqrt{\frac{1}{S}\sum_{i=1}^{S}|x_i|^2}}, \quad (6)$$

where $S$ is the length of the channel input and output for training, $x_i$ represents the input data without normalization, and $\tilde{x}_i$ is the data after normalization. The fiber channel input data is relatively small, which can slow down the convergence of the model and reduce the accuracy of the trained model. Therefore, the unit power normalization is very important to control the average absolute value of the input around 1. To ensure arbitrary data satisfying the $p_{data}$ can be generated from the noise vector, the length of the noise vector is required to be larger than the generative data dimension, so we set the length of $z$ to 10. In this letter, the labels for real data and fake data are selected randomly from [0.7~1.2] and [0~0.3], respectively, instead of fixed values, 0 and 1, which allow the existence of noise in the discriminator but enhance the generator ability [26]. Adam optimization algorithm is used to update the parameters of GAN [27]. The weights of generator and discriminator are initialized by He initialization as described in [28]. This methodology helps with the convergence of deep models with ReLu-like activation functions. And all the biases are initialized to 0. The learning rates of these two neural networks are both set to 0.0005 and the batch size is set to 500.

## IV. RESULTS AND DISCUSSION

In the simulation, we keep dispersion and nonlinearity coefficient as a constant but change the transmission distance and the launch optical power to control the dispersion and nonlinear intensity. For convenience, we use a pair of numbers ($D, P$) to represent the transmission conditions, where $D$ represents the transmission distance and $P$ represents the optical launch power. In our simulation, all the transmission distances consist of multiple spans and each span length is 50 km. For example, a 200-km link equals to the 4-spans × 50-km fiber lines. The results and the demonstrations are presented from multiple dimensions, including constellations, optical waveforms, spectra, and the normalized MSEs of SSFM and GAN output. Constellations after CD and DBP compensation are scattered to present the characteristics of dispersion and nonlinearity. Meanwhile, the accuracy of the channel transfer function modeled by GAN can be verified from the shape of constellations after DBP compensation, as described in Section II. Optical waveforms and spectra of channel output are plotted to demonstrate the accuracy of time- and frequency-domain characteristics. The normalized MSEs of SSFM and GAN output are carried out to compare the gap of SSFM and GAN output quantitatively.

To model links of different transmission conditions, we collected channel input and output of different link lengths and optical launch powers. Our training dataset size is set to $1 \cdot 10^6$ for most transmission conditions. And the size of training dataset is different for different transmission conditions. For longer fiber lines and bigger launch powers, the more training data are required. The number of training data for specific conditions and the detailed explanations are presented at the following results.

### A. The channel modeling capability of GAN

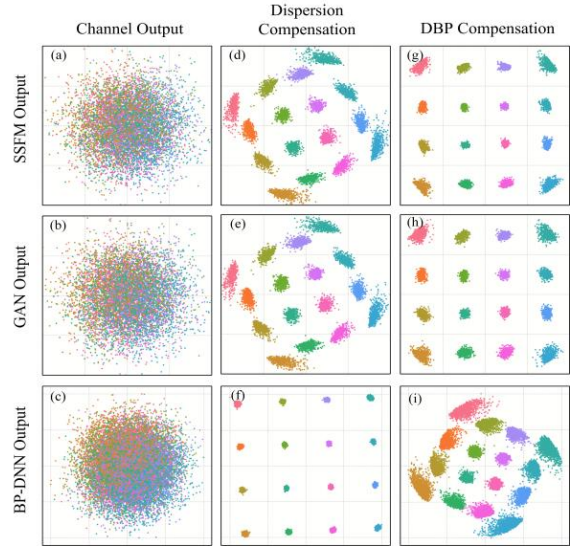

Fig. 7. Constellations of SSFM, GAN and BP-DNN output with/without compensation at (50 km, 10 dBm): (a), (b), (c) channel output, (d), (e), (f) channel output after CD compensation, (g), (h), (i) channel output after DBP compensation.

We firstly demonstrate the fiber channel modeling ability of GAN. As a comparison, BP-DNN with the same structure and input of generator is also utilized to model the fiber channel, but the optimization aim is to minimize the MSE between the output of SSFM and BP-DNN. Fig. 7 presents the constellations of output data generated by different models with or without compensation. The transmission distance is 50 km, and the input fiber power is 10 dBm, represented by (50 km, 10 dBm) for convenience. Fig. 7(a) to Fig. 7(c) are the channel output modeled by SSFM, GAN, and BP-DNN on testing data, respectively. Then CD compensation is utilized to eliminate ISI so that the characteristics of SPM nonlinearity are reflected in the constellation [29]. Fig. 7(d) illustrates the constellation of SSFM output after CD compensation, where the symbols have a common phase rotation. Note that symbols with different amplitudes lead to different phase rotations [30]. With the constellation of GAN output after CD compensation, Fig. 7(e) shows that the ISI of GAN output is successfully equalized. The same constellations changes indicate that GAN has learned accurate CD characteristics. If we compare Fig. 7(e) with Fig. 7(d), we observe that the nonlinear phase rotations of GAN output are identical with SSFM output, which proves GAN has also learned SPM nonlinearity characteristics of the fiber channel properly. We also utilize the DBP compensation, an inverse process of channel transmission, to restore the channel output close to the original channel input. As can be seen in Fig. 7(g), CD-induced ISI and nonlinearity-induced phased rotations are eliminated after DBP compensation, and the constellation is restored to the 16QAM constellation. Similarly, GAN output is also recovered to the original 16QAM transmitted data after DBP compensation, as shown in Fig. 7(h). The accurate constellation changes indicate that the distribution generated by GAN is equivalent to the channel transfer function simulated by SSFM.

As for BP-DNN, the generative output can be recovered by CD compensation, but it ignores the nonlinear phase rotation, as shown in Fig. 7(f). It is because BP-DNN uses MSE as the loss function, which aims to fit data with the global average optimization by ignoring the specific features of signals [31]. Specifically, the phase rotation with noise will increase the MSE and is hard to fit by an average method. Consequently, BP-DNN abandons the modeling of phase rotation and converges to the mean value of QAM data. If the BP-DNN output is equalized by DBP, ISI of the signal caused by CD can be eliminated, but the inverse phase rotation exists, as shown in Fig. 7(i).

The loss of generator and discriminator during training for the modeling of (50 km, 10 dBm) transmission is presented as Fig. 8. The epoch number is set to 20 and the batch number of each epoch is 2000. The total points of the loss are 40000 ($20\times2000$). The losses of both generator and discriminator decrease rapidly, and then converge to a certain value. The converge values of generator and discriminator are around 0.66 and 1.37, respectively. The losses then fluctuate in a small range around the converge value. The characteristics of GAN loss correspond to the working mechanism of GAN, i.e., the generator and discriminator are trained in an adversarial progress, and aim to find a Nash equilibrium. We find that the larger fluctuation range is also acceptable for the channel modeling of other transmission conditions.

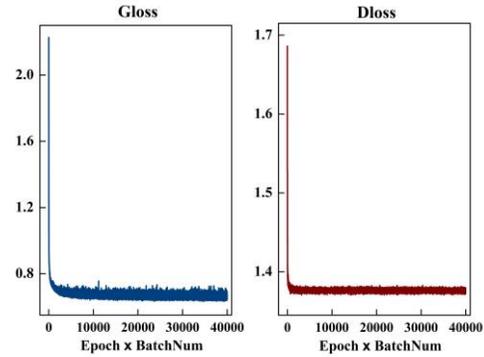

Fig. 8. The loss of generator and discriminator for the channel modeling of GAN at (50 km, 10 dBm) transmission. Notation Gloss and Dloss represent the losses of generation and discrimination, respectively.

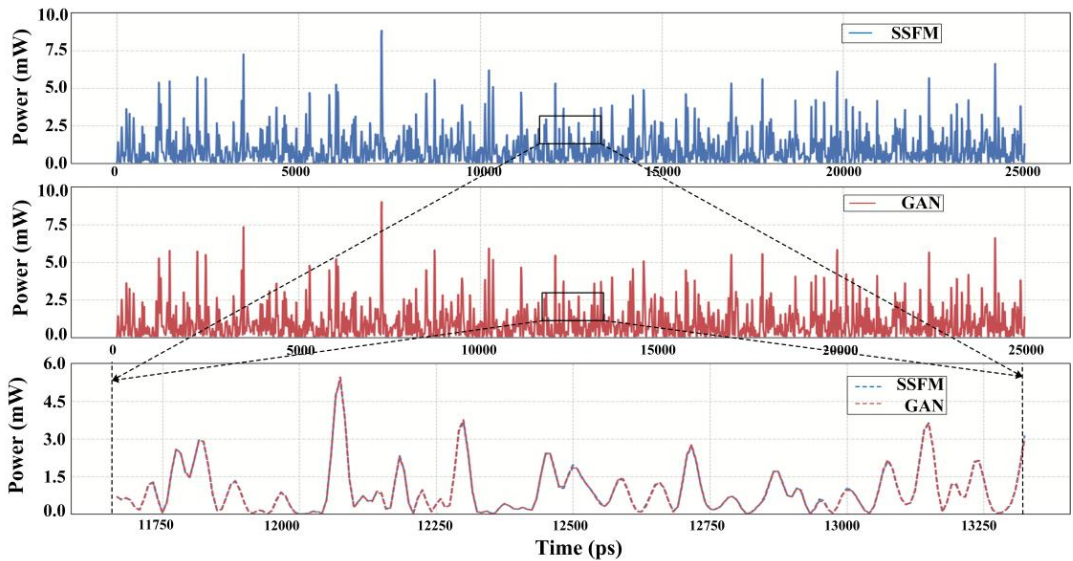

Fig. 9. Amplitudes of optical waveforms of channel output based on SSFM and GAN at (200 km, 0 dBm).

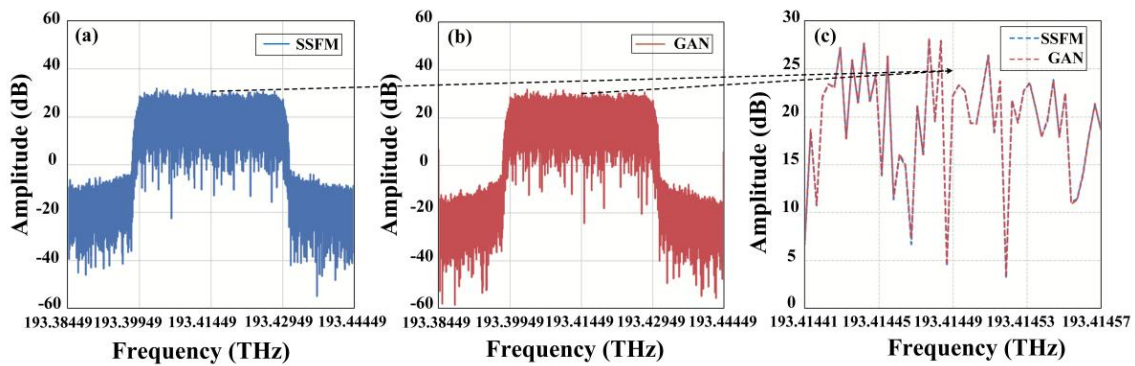

Fig. 10. Optical spectra of channel output based on SSFM and GAN at (200 km, 0 dBm).

For a more diverse and accurate demonstration, optical waveforms and spectra of optical fiber channel output simulated by SSFM and GAN at the transmission of (200 km, 0 dBm) are also presented in Fig. 9 and Fig. 10. Fig. 9 illustrates the amplitudes of optical waveforms of SSFM-based channel output and GAN-based channel output. From the overall view of the optical waveform, the channel output generated by SSFM and GAN have similar powers at the same time. We zoom in the waveforms of one section and find that the waveform of GAN output and SSFM output are overlapped to a great extent. Fig. 10 displays the optical spectra of SSFM-based channel output and GAN-based channel output within 30 GHz, which is the same as the transmission baud rate, to show the signal characteristics. The spectra of GAN and SSFM output are similar on the whole, and the enlarged spectra near the central frequency band show the high-overlap ratio. The high degree of consistency for optical waveforms and spectra demonstrates that the time- and frequency-domain characteristics of signals are modeled accurately by GAN.

To express the gap between GAN-based model and the SSFM-based model quantitatively, we adopt the normalized MSE of power as the method in [6], which is expressed as

$$MSE\_nor = \frac{\sum_1^m (\hat{y}-y)^2}{\sum_1^m y^2}, \quad (7)$$

where $m$ is the data length, $y$ is the SSFM output data, and $\hat{y}$ is the generated data by GAN. As described in [6], the acceptable upper bound of normalized MSE is set to 0.02. Fig. 11 presents the normalized MSEs of signal amplitudes in the time domain at different transmission distances. Note that the training dataset size is $4 \cdot 10^6$ for the modeling of the transmission at (500 km, 0 dBm) and (1000 km, 0 dBm). The value of MSE increases as the fiber length increases since the joint effect of high-intensity dispersion and nonlinearity. After 20 iterations on the test dataset, the mean value of normalized MSE at (1000 km, 0 dBm) is 0.00925, much lower than 0.02, which demonstrates the capabilities of GAN for the long-haul transmission. Theoretically, GAN can be used to model the channel with longer transmission distance by increasing the length of the condition vector.

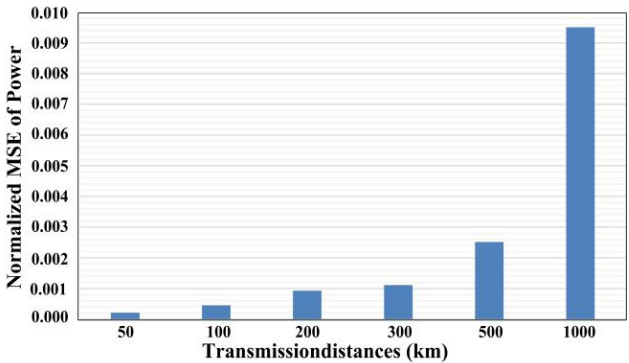

Fig. 11. Normalized MSEs of signal amplitudes in time domain at different transmission distances.

The above results show the ability of GAN to model the long-haul transmission at low nonlinear intensities. We also compare the waveforms and spectra generated by the GAN with that generated by the SSFM in a highly nonlinear region. The channel modeling at the condition of (50 km, 10 dBm) and (100 km, 10 dBm) is taken as an example. The number of raining data for the transmission at (100 km, 10 dBm) is $2 \cdot 10^6$. Considering the large noise induced by the nonlinearity, the optical signals are matched by a RRC filter to remove the noise out of the signal band. The waveforms and spectra are shown as Fig. 12 and Fig. 13. The generated waveforms and spectra by GAN approximate to that generated by SSFM. We also calculated the normalized MSEs to compare the results quantitatively. The MSEs of waveforms in time domain at (50 km, 10 dBm) and (100 km, 10 dBm) are 0.00072 and 0.00296, respectively. The normalized MSEs of spectra are 0.00028 and 0.00072. The MSEs are far lower than the upper bound 0.02, which demonstrates the ability of GAN to model the fiber channel in the highly nonlinear region.

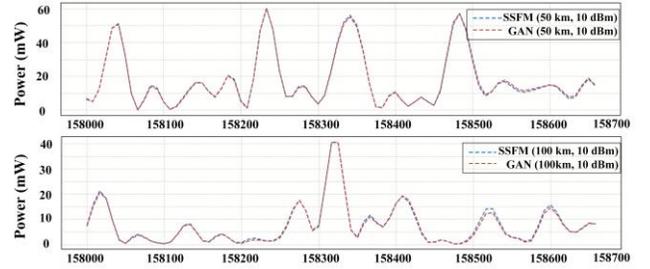

Fig. 12. Amplitudes of optical waveforms of channel output based on SSFM and GAN at (50 km, 10 dBm) and (100 km, 10 dBm).

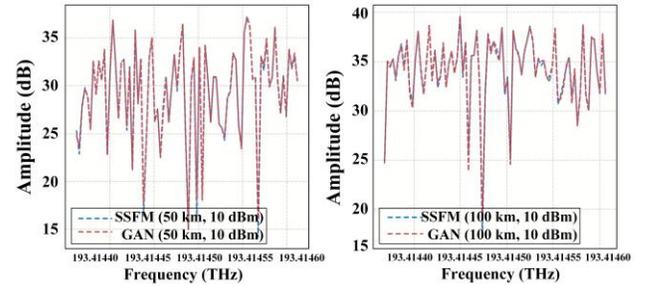

Fig. 13. Optical spectra of channel output based on SSFM and GAN at (50 km, 10 dBm) and (100 km, 10 dBm).

Optical fiber transmissions with different optical launch powers are also investigated. The power of training data for a 200-km fiber channel consists of (-2, 0, 2, 4 dBm). Fig. 14 illustrates the constellations of channel output based on SSFM and GAN after DBP compensation for various powers. The normalized MSEs between the output of GAN and SSFM for the 200-km fiber links launched with (-2, 0, 2, 4 dBm) are 0.0009, 0.000925, 0.000979, and 0.0018, respectively. The normalized MSEs increase with the launch powers since the distortions caused by the high-intensity nonlinearity. But the MSEs are still much lower than the acceptable upper bound 0.02. The remarkable similarities and low normalized MSEs between GAN output and SSFM output illustrate the high ability of GAN to model the optical fiber channel of various launch powers simultaneously.

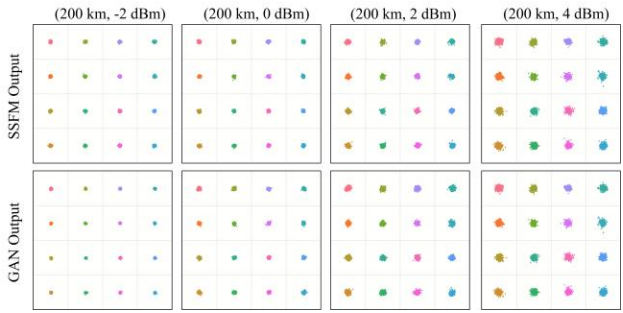

Fig. 14. Constellations of SSFM output and GAN output after DBP compensation at (-2, 0, 2, 4 dBm), which are contained in the training dataset.

### B. Generalization of GAN

The generalization of GAN is essential for the practical applications of channel modeling, especially for different transmission conditions and the various input. The GAN-based channel model we propose can adjust the input fiber power flexibly. As demonstrated above, the GAN-based fiber model possesses the same results with the SSFM-based model for a 200-km fiber channel at (-2, 0, 2, 4 dBm). To investigate the generalization of GAN for other launch powers, we test the data at (-1, 1, 3, 4.5 dBm), which are never occurred during the training process. From the constellations results of SSFM output and GAN output after DBP compensation, as shown in Fig. 15, we observe that GAN learns different nonlinearity intensity corresponding to different launch optical power successfully by one generator. The normalized MSEs of the transmission at (-1, 1, 3, 4.5 dBm) are 0.000910, 0.000951, 0.00136, and 0.0021, respectively. The trend of these MSEs obtained from the testing dataset has the consistence with that from the training dataset. Therefore, the results illustrate the generalization of GAN under different optical launch powers.

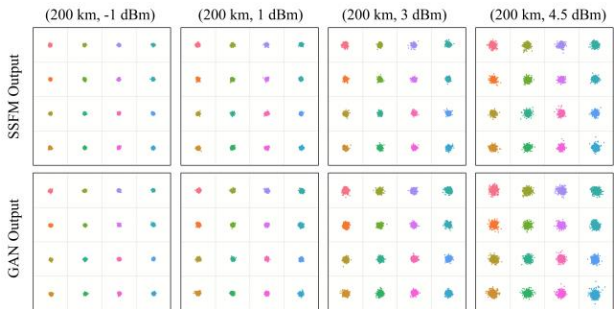

Fig. 15. Constellations of SSFM output and GAN output after DBP compensation at (-1, 1, 3, 4.5 dBm), which are not contained in the training dataset.

During the training process, the training dataset contains only 16QAM symbols with uniform distribution. On the test dataset, we study the channel input data of different modulation formats, including QPSK, 32APSK, and 64QAM. As shown in Fig. 16, the output of SSFM and GAN after DBP compensation show the good generalization ability of GAN for modulation formats, whether high-order or low-order. Moreover, we add Gaussian noise to the input 16QAM data to change the uniform distribution of channel input to the Gaussian distribution. The results show that GAN is also applicable to model Gaussian distribution data. We also calculate the normalized MSEs of channel output matched with RRC filter corresponding to the channel input with these different modulation formats and distributions. The MSEs are 0.001564, 0.001069, 0.001064, and 0.000718, respectively. The MSEs of input with QPSK, 32APSK, and 64QAM are slightly higher than that of 16QAM at the same conditions, but are still very low. The MSE of input with Gaussian distribution is similar with the results of the uniform distribution. Therefore, we believe that GAN has good generalization ability for the input with different modulation formats and distributions.

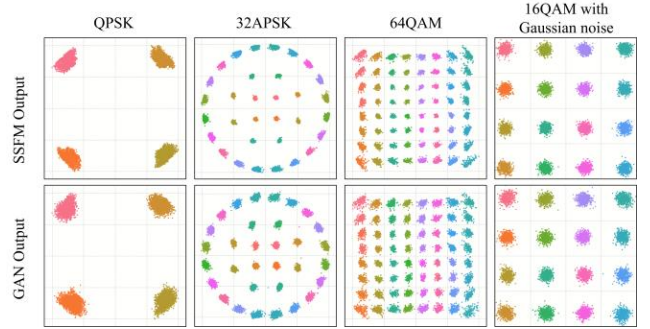

Fig. 16. Constellations of SSFM output and GAN output after DBP compensation with different modulations and distribution at (50 km, 10

### C. Complexity analysis

To compare the complexity of the SSFM-based model and the GAN-based model, we theoretically calculate the multiplication computation required in the modeling process, which is commonly used to compare the hardware complexity of the algorithms [32]. The principal calculation amount of SSFM is FFT operation, and each transmission step contains 2 FFT operations [33]. Each FFT operation contains $2N\log_2 N$ real number multiplication operations, i.e. $C_{FFT}=2N\log_2 N$, where $N$ is the FFT size [34]. Taking all transmission steps into account, the multiplication amount of SSFM can be expressed as

$$C_{SSFM} = N_{span}N_{step}(4N\log_2 N + kN), \quad (8)$$

where $N_{span}$ is the number of spans, $N_{step}$ is the number of steps per span based on SSFM and $k$ presents the rest of the multiplication except FFT in one step. The exponential operation of each step is ignored. It can be concluded that the complexity increases linearly with fiber transmission distance.

In the case of NN, the output of the $jth$ neuron in the $lth$ layer can be expressed as

$$a_j^l = \sigma(\sum_k w_{jk}^l a_k^{l-1} + b_j^l), \quad (9)$$

where $\sigma$ represents the activation function, $w^l$ represents a weight matrix for the $lth$ layer, and $w_{jk}^l$ is the weight in the $jth$ row and $kth$ column of the weight matrix [35]. The sum of Eq. (9) is overall neurons in the $(l-1)th$ layer. Therefore, the total multiplication of the $lth$ layer is $n_{l-1} \times n_l$, where $n_l$ is the all neurons in the lth layer. For a fully-connected neural network (FCNN) with $M$ layers, it needs $(n_1 n_2 + n_2 n_3 + \ldots + n_{M-1} n_M)$ multiplications, neglecting the activation function. In this work, $n_1$ is set to increase with the distance and is set to 80 for each span length, as introduced in Section III. Thus, the dimension of the first layer can be expressed as $n_1 = 80 N_{span} + D_z + 8$, where $D_z$ is the length of latent code and 8 represents the number of generator output each time. The dimensions of rest layers are kept constant for arbitrary distance. The layer number of the generator is set to four in this work and considering the generator generates four samples each time, the multiplication number of GAN can be expressed as

$$C_{GAN} = N/4[(80N_{span} + D_z + 8)n_2 + n_2n_3 + n_3n_4 + n_4n_5]. \quad (10)$$

It can be seen that the complexity of GAN is also linear with the transmission distance.

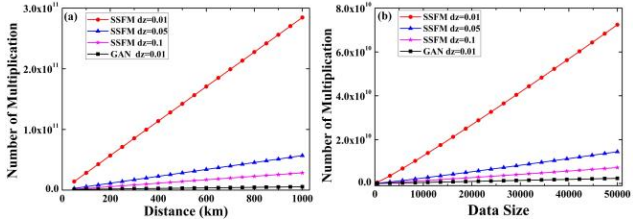

Fig. 17. Number of multiplications vs. (a) distance and (b) data size of channel model based on SSFM and GAN. The notation dz represents the step distance.

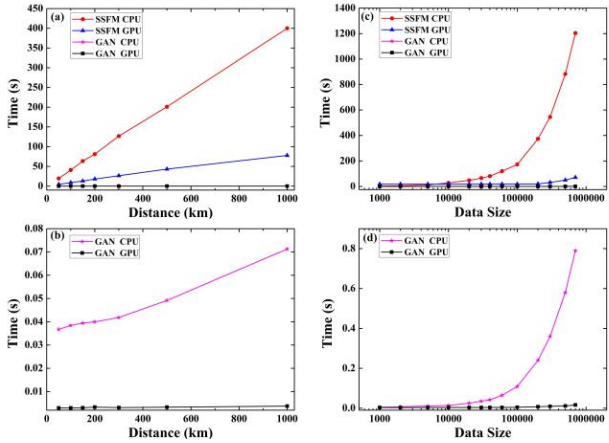

Fig. 18. Running time vs. distance and data size of channel model based on SSFM and GAN. (b) and (d) are the enlarged figure of (a) and (c) to show the detail results of GAN.

With all the parameters taken in, the relationship between the total multiplication computation and the distance is plotted as shown in Fig. 17(a), where the data length is set to 40000. The complexity decreases as the step distance increases. The small step distance ensures the accuracy of the fiber channel model. In this work, GAN is utilized to model the fiber channel simulated by SSFM with 0.01 km per step. The number of multiplications of SSFM at 1000-km fiber length and 0.01-km step distance is about $2.85 \cdot 10^{11}$, whereas the multiplications of GAN is just about $5.57 \cdot 10^9$, which means the complexity of GAN-based model is only about 2% of the SSFM-based model at 1000-km fiber transmission. In addition to the transmission distance, the computational complexity also relates to the length of the transmitted data. Therefore, the relationship between the number of multiplications and the transmitted data size is also be explored, as shown in the Fig. 17(b), where the data size is set to from 500 to 50000, and the transmission distance is set to 200 km with 0.01-km step distance. With the increase of data size, the complexity of SSFM-based and GAN-based model both increases. When the data length is 500, the number of multiplications of SSFM is about $4.59 \cdot 10^8$, whereas the multiplications of GAN is just about $2.35 \cdot 10^7$. The results show that the complexity of GAN-based model is always lower than that of SSFM-based, even for the short data length.

We also record the running time of the SSFM-based model and the GAN-based model at the same hardware and software conditions. The data length is also set to 40000. The codes of these two models run on the same server with an NVIDIA GeForce RTX 2080Ti Computer Graphics Cards. As shown in Fig. 18(a), the time of the SSFM-based model increases with the distance on the central processing unit (CPU) and GPU, and due to the parallel computing, the running time on GPU is reduced to a large extent. For the 1000-km transmission, SSFM takes 400 seconds on CPU and 78 seconds on GPU. The time of GAN on CPU grows with distance, and the running time is about 0.07 seconds at 1000-km fiber transmission as shown in Fig. 18(b). It is worth noting that the running time of GAN on GPU remains constant around 0.0037 seconds, because the amount of computation caused by the increase of the distance is calculated in parallel by GPU. It takes less than 0.1 seconds for GAN no matter on GPU or CPU, which is significantly shorter than SSFM. The executed time of SSFM-based and GAN-based model for different data size is also recoded, as shown in Fig. 18(c) and Fig. 18(d). The data size is set from 1000 to 7000000. The executed time of SSFM-based model and GAN-based model on CPU increases with the increases of data size. The time of the model based on SSFM and GAN on GPU remains constant under the data size of 100000 and then increases with the data size. When the data length is 7000000, the executed time of GAN is less than 0.1 seconds, which is far shorter than the executed time of SSFM, 1204 seconds on CPU and 71 seconds on GPU. The results show the running time of GAN on the software can be significantly reduced compared with SSFM. In addition, as long as the training is finished, the generator of GAN is an FCNN, which is the simplest structure of the deep learning models.

D. *Modified loss function improvement*

As described in Section III, we add reconstruction loss, Ly, to the loss function of the generator, which can address the problem of inaccurate amplitudes of GAN. Fig. 19 illustrates optical spectra of channel output generated by SSFM, GAN without and with Ly loss, respectively. Without Ly loss, the spectrum of the channel output generated by GAN has a large impulse at the center carrier frequency and three large impulses at the frequency of 30 GHz and 60 GHz away from the center carrier frequency. The central impulse means GAN is insensitive to signal amplitudes. In addition, the GAN generates four samples, i.e., one symbol, every time. Thus, the cyclic output of GAN generates periodic samples resulting in impulse at sideband. After adding Ly loss, the generated data are limited to the real amplitude range of the channel output, which can relieve the periodicity and improve the accuracy of generated data. As presented in Fig. 19 (c), the pike at the center carrier frequency disappears, and the other three pikes are significantly reduced from 50 dB to 5 dB.

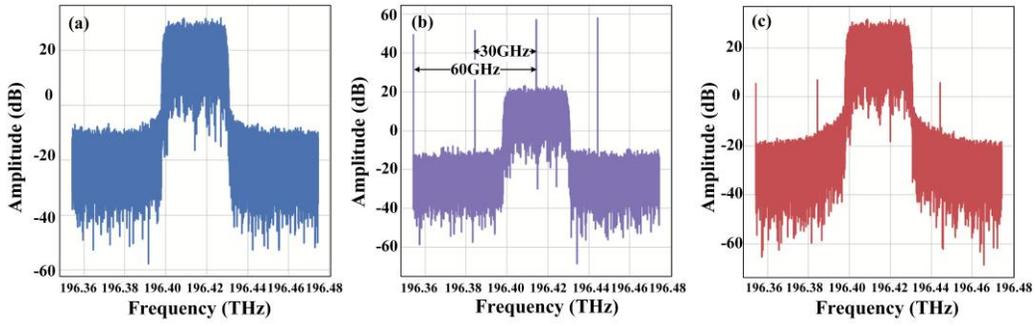

Fig. 19. Optical spectra of channel output (200 km, 0 dBm): (a) SSFM-based channel output (b) GAN-based channel output without Ly loss, (c) GAN-based channel output with Ly loss.

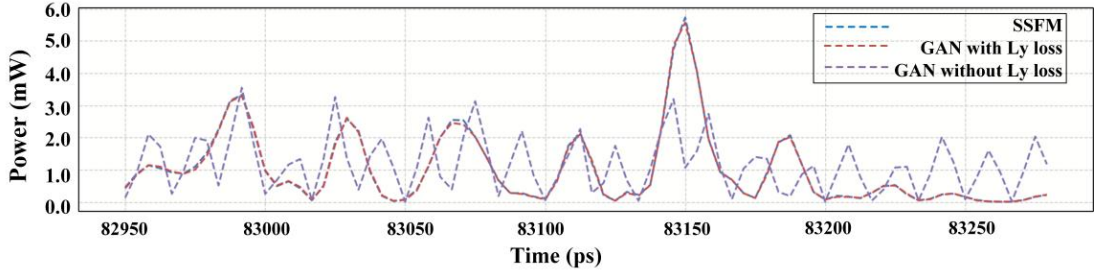

Fig. 20. Amplitudes of optical waveforms of channel output modeled by SSFM, GAN without Ly loss, and GAN with Ly loss at (200 km, 0 dBm).

From the aspect of the time domain in Fig. 20, the waveform of GAN output without Ly loss has many pikes, i.e. the amplitudes of the signal are constantly shaking. This phenomenon is called jitters in this article. These jitters reflect the inaccurate generation of GAN compared with the real output generated by SSFM. After using Ly loss as the regularization term of the generator loss function, the amplitudes of GAN output are restricted to approximate the amplitudes of the SSFM output during the training process. As shown in Fig. 20, the jitters of waveforms are suppressed and GAN realizes the accurate generation with Ly loss.

*E. Mode collapse of GAN*

During our simulation, we find that the characteristics of dispersion can be easily learned, but the nonlinearity modeling and the long-haul transmission modeling encountered difficulties. As presented in Fig. 21, compared with the SSFM output after CD compensation in the red circle, GAN output has the same common phase rotation $\theta$ with the SSFM output but ignores the differences between the data with different amplitudes. Then the GAN output cannot be recovered to the original data by DBP compensation, inducing some noise of the signal. Increasing the epoch number is useless to solve this problem. This phenomenon is caused by the mode collapse of GAN, where some of the nonlinear features of transmitted data are represented, but the others are ignored. As the distance increases, the dimension of the input condition vector $x$ increases, which means that the channel transfer function $p(y|x)$ becomes more complicated for modeling. Meanwhile, the large nonlinearity leads to the huge distortions, which also leads to the difficulties for the modeling of the fiber channel.

In the simulation, we increase the training dataset size from $2·10^6$ to $4·10^6$ for the modeling of (500 km, 0 dBm) link. The large training dataset enriches the diversities of training samples and enhances the learning abilities of generator. After increasing the training dataset size, the phase rotation related to the signal amplitudes are learned by GAN successfully. The results shown in Fig. 21 prove that the mode collapse of GAN can be relieved by increasing the amount of training data. In the simulation, the appropriate training dataset size can be selected according to different transmission conditions. Fortunately, in the field of communication, the amount of channel data is infinite by collecting data from the transmission link continuously, which ensures the dataset size required for GAN in the channel modeling.

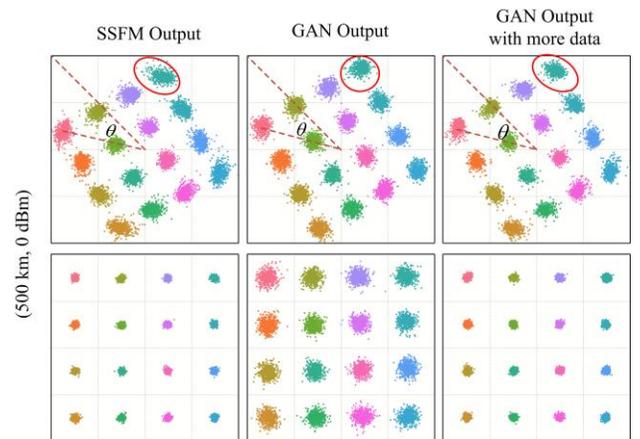

Fig. 21. Constellations of GAN output after CD compensation and DBP compensation at (500 km, 0 dBm).

## V. Conclusion

This paper demonstrates that adversarial learning with conditional GAN is a good candidate to model optical fiber channel with much lower calculation complexity and shorter running time. The multiplication number is only 2% of SSFM at 1000-km fiber transmission. The normalized MSE is 0.00925 at (1000 km, 0 dBm), which is much lower than the upper bound of the normalized MSE. GAN also has good generalization abilities for launch powers, modulation formats, and data distributions, which provides flexibilities and versatility for the application of channel modeling. GAN also has the capabilities to model wavelength- and polarization-division multiplexed optical channel, because the structure of GAN proposed in this article has no limitation on the input signals. The new condition vector structure and parameters of GAN may be required since the complicated inter channel interferences, which would be further investigated in the future. GAN also has the potential to model more devices and even the whole physical optical communication system. As a low complexity and DL-based modeling tool, GAN is compatible to be embedded in other neural networks for better signal design, such as new modulation format, channel coding, and shaping filter.